\documentstyle[12pt,epsf]{article}
\newcommand{\sect}[1]{\setcounter{equation}{0}\section{#1}}

\def\be{\begin{equation}}
\def\ee{\end{equation}}
\def\ba{\begin{eqnarray}}
\def\ea{\end{eqnarray}}
\def\bq{\begin{quote}}
\def\eq{\end{quote}}

\def\PL{{ \it Phys. Lett.} }
\def\PRL{{\it Phys. Rev. Lett.} }

\def\PR{{\it Phys. Rev.} }
\def\MPL{{\it Mod. Phys. Lett.} }

\def\CQG{{\it Class. Quant. Grav.} }

\begin{document}
\thispagestyle{empty}
\begin{flushright}
Stanford SU-ITP-98-143
%\\ hep-th/9807123
\\ July 1998
\end{flushright}
\vspace*{1cm}
\begin{center}
{\Large \bf Hawking radiation by effective two-dimensional theories}\\
%{\Large \bf 4d
%minimally coupled scalar fields }\\
%{\Large \bf and Hawking radiation}\\
\vspace*{2cm}
R. Balbinot\footnote{E-mail: balbinot@bologna.infn.it}\\
\vspace*{0.2cm}
{\it Dipartimento di Fisica dell' Universit\`a di Bologna}\\
{\it and INFN sezione di Bologna}\\
{\it Via Irnerio 46, 40126 Bologna, Italy  }\\
\vspace*{0.5cm}
and A. Fabbri\footnote{Supported by an INFN fellowship. E-mail:
afabbri1@leland.stanford.edu}\\
\vspace*{0.2cm}
{\it Department of Physics, Stanford University}\\
{\it Stanford, CA 94305-4060, Usa}\\
\vspace*{3cm}
{ \bf Abstract}
\end{center}
Recently proposed 2D anomaly induced effective actions for the matter-gravity 
system are critically reviewed. Their failure to correctly reproduce 
Hawking's black hole radiation or the stability of Minkowski space-time
led us to a modification of the relevant ``quantum'' matter
stress energy tensor that allows physically meaningful results to be 
extracted.

\vfill
\setcounter{page}{0}
\setcounter{footnote}{0}
\newpage
\sect{Introduction}
Hawking's remarkable discovery \cite{Hawking} that black holes emit quantum thermal radiation 
at a temperature inversely proportional to their mass (i.e. 
$T_H= \frac{1}{8\pi M}$ in units where $\hbar=c=G=k_B=1$) triggered, in the mid 
seventies, a large scale investigation of quantum effects in strong 
gravitational  fields (see for example \cite{BirDav}).
The framework used was quantum field theory in curved 
space, a semiclassical approach in which only the matter fields are quantized, 
whereas gravity is still described classically according to General 
Relativity.  Its dynamical evolution is driven by the expectation value of 
the renormalized energy-momentum tensor operator of the quantized matter 
fields, i.e. $\langle T_{\mu\nu}\rangle$, according to the semiclassical
Einstein equations 
\be
G_{\mu\nu} (g_{\mu\nu})=8\pi \langle T_{\mu\nu}(g_{\mu\nu})\rangle \ .
\label{i}
\ee
The left-hand side  (l.h.s.) is the Einstein tensor for the space-time metric 
$g_{\mu\nu}$, while the r.h.s. represents the expectation value of the
stress tensor of the matter fields propagating on that space-time. \\
According to Wald's axioms \cite{Wald}
$\langle T_{\mu\nu}\rangle $ 
must be conserved, $\nabla_{\mu}\langle T^{\mu}_{\nu}\rangle=0$ ,
and vanishing for Minkowski space-time, so that eq. (\ref{i}) can 
make sense. One further important thing to note is the presence in 
$\langle T_{\mu\nu}\rangle$ of a trace anomaly (see 
for instance \cite{Duff}). 
For conformally invariant 
fields the expectation value of the trace 
 $\langle T^{\alpha}_{\alpha}\rangle$ 
is nonzero, unlike its classical counterpart, and independent of the state 
in which the expectation value is taken. It is  
completely 
expressed in terms of geometrical objects as 
\be 
\langle T^{\alpha}_{\alpha}\rangle = (2880\pi^2)^{-1}\{a
C_{\alpha\beta\gamma\delta}C^{\alpha\beta\gamma\delta} +
b(R_{\alpha\beta}R^{\alpha\beta}- \frac{1}{3}R^2)+
c\Box R + dR^2 \}\ ,
\label{tial}
\ee
where the coefficients in front of each of the above geometrical 
tensors are known and depend on the spin of the quantum field 
under consideration \cite{BirDav}. 
These features will be of fundamental importance 
throughout the paper.  \\
Technically, one has to construct  $\langle T_{\mu\nu}(g_{\mu\nu})\rangle$
for a sufficiently large class of metrics (according, for example, 
to the symmetry of the problem) and then solve eq. (\ref{i}) 
self consistently for the metric. Unfortunately, general expressions 
for  $\langle T_{\mu\nu}(g_{\mu\nu})\rangle$ are not available, except when 
the degree of symmetry of the problem is sufficiently high, for instance 
conformally invariant fields in homogeneous and isotropic 
space-times \cite{Anderson}
where that the trace anomaly determines completely $\langle T_{\mu\nu}\rangle$.
This is not the case, however,  for black holes. 
 $\langle T_{\mu\nu}(g_{\mu\nu})\rangle$ for a sufficiently arbitrary 
(even spherically symmetric) black hole spacetime is not known even 
approximately. So the evolution of the black hole as it Hawking emits
(the so called backreaction) is an open problem. \\
Much effort has been devoted to understand all the features of 
$\langle T_{\mu\nu}\rangle$  for the Schwarzschild black hole geometry 
in order to get some insight in the backreaction. Note, however, that the 
Schwarzschild spacetime is not a solution to eq. (\ref{i}) since the l.h.s.
vanishes, unlike the r.h.s. . 
\par \noindent
Using analytical methods (which can be improved numerically) 
one can find reasonable approximations of $\langle T_{\mu\nu}\rangle$
for various kinds of quantum fields propagating on the 
Schwarzschild space-time ( see for example \cite{CanOHi}). 
Within this context,
three quantum states might be proposed as a suitable candidate for the
vacuum:
\par \noindent
i) the Boulware state $|B\rangle$ \cite{Boulware}, 
defined by requiring normal modes
to be positive frequency with respect to the Killing vector 
$\frac{\partial}{\partial t}$, according to which the region exterior 
to the horizon is static. The stress tensor in this zero temperature 
state describes the vacuum polarization outside a static star whose
radius is bigger than the Schwarzschild one (i.e. $r>2M$).
As $r\to\infty$ $\langle B|T_{\mu\nu}|B\rangle \to 0$. 
The Boulware state corresponds to our familiar concept of an empty
state for large radii. Symbolically, $|B\rangle \to |M\rangle$, where
$|M\rangle$ is Minkowski vacuum. However, $|B\rangle$ is pathological
at the horizon as it diverges when evaluated in a free falling frame.
\par \noindent
ii) the Hartle-Hawking state \cite{HHI}, 
defined by taking incoming modes to be positive
 frequency with respect to the canonical affine parameter on the future
horizon (Kruskal coordinate V) and outgoing modes to be positive frequency 
with respect to the canonical affine parameter on the past horizon 
(Kruskal's U). $\langle H|T_{\mu\nu}|H\rangle $ is well behaved on both
future and past horizons. This state is not empty at infinity, corresponding
 to a thermal distribution of quanta at the Hawking temperature $T_H=
\frac{1}{8\pi M}$, i.e. 
\be
\langle H| T^{\mu}_{\nu}   |H\rangle \sim  \frac{1}{2\pi^2}
\int_0^{\infty} \frac{w^2 dw}{e^{8\pi M w} -1}
\pmatrix{ -1 & 0 & 0 & 0 
\cr 0 & 1/3 & 0 & 0 
\cr 0 & 0 & 1/3 & 0 
\cr 0 & 0 & 0 & 1/3 \cr}\ . 
\label{lungo}
\ee
That is, the state $|H\rangle$ corresponds to a black hole in 
equilibrium with an infinite reservoir of black body radiation .
\par \noindent   
iii) the Unruh state $|U\rangle$ \cite{Unruh}, 
defined by taking modes that are incoming
from past null infinity to be positive frequency with respect to 
$\frac{\partial}{\partial t}$, while those that emanate from the past horizon 
to be positive frequency with respect to $U$. At infinity this state 
corresponds to an outgoing flux of blackbody radiation at the black 
hole temperature $T_H$
\be
\langle U| T^{\mu}_{\nu}   |U\rangle \sim \frac{L}{4\pi r^2}
\pmatrix{ -1 & 1 & 0 & 0 
\cr 1 & 1 & 0 & 0 
\cr 0 & 0 & 0 & 0 
\cr 0 & 0 & 0 & 0 \cr}, 
\label {un}
\ee
where $L$ is the luminosity factor of the hole.
$\langle U|T_{\mu\nu}|U\rangle$ is regular, in a free falling frame, 
on the future horizon but not on the past horizon. As $r\to 2M$ 
the $(r,t)$ part reads 
\be
\langle U| T^{\mu}_{\nu}   |U\rangle \sim \frac{L}{4\pi }
  \pmatrix{ (1-2M/r)^{-1} & -r^2 \\  
\cr r^2 (1-2M/r)^{-2} & -(1-2M/r)^{-1} \cr} \ .
\label{buuu}
\ee
The state $|U\rangle$ is supposed to best approximate the state of the 
quantum fields outside a collapsing star as its surface approaches the
horizon. This implies that the divergence on the past horizon is spurious
since this portion of the Schwarzschild spacetime is not physical being covered
by the collapsing body. 
\par \noindent
Starting from these results attempts have been made to solve at least 
perturbatively the backreaction for a black hole enclosed in 
a box \cite{York}. 
The approach followed (Hartree-Fock like) is to write the backreaction 
equations  (\ref{i})  as follows 
\be
G_{\mu\nu} (g^s_{\mu\nu}+\delta g_{\mu\nu} ) 
= 8\pi \langle H| T_{\mu\nu}(g_{\mu\nu}^s)|H\rangle ,
\label{basc}
\ee
where $g_{\mu\nu}^s$ represents the Schwarzschild metric, and 
one solves eqs.  (\ref{basc}) linearizing in the static spherically symmetric
perturbation $\delta g_{\mu\nu}$. A similar approach is much more difficult to
implement for an evaporating black hole. Attempts have been made by modelling 
the time dependent geometry near the horizon (and also asymptotically) 
by a Vaidya space-time and some insights in the evaporation process can be 
extracted \cite{Balb}. \\
The main difficulty to attack the backreaction equations 
 (\ref{i}) is, as we already pointed out, 
the absence of an explicit expression of 
$\langle T_{\mu\nu} \rangle$ for a sufficiently general (for example 
spherically symmetric) evaporating black hole geometry.    
This fact can be circumvented in two space-time dimensions. For conformally
invariant and minimally coupled scalar
fields one is able to obtain an expression for the 2D stress tensor
$\langle T_{ab}(g_{ab})\rangle$ for a generic 2D metric $g_{ab}$ \cite{BirDav}.
This expression can be formally obtained starting from the well known 
Polyakov action \cite{Polya}. Explicit evaluation of $\langle T_{ab}\rangle$ 
for a 2D Schwarzschild geometry in the state $|B\rangle$, $|H\rangle$
and $|U\rangle$ gives results which are in good qualitative agreement
with the 4D $ \langle T_{\mu\nu}\rangle $
described earlier. \\
This nice agreement and the possibility of having a $\langle T_{ab}\rangle$ 
for an arbitrary 2D metric has triggered extensive investigation of 
2D versions of the backreaction equations (\ref{i}) in the hope of learning 
something about physical (i.e. 4D) black hole evaporation. In all 
such 2D models 
gravity (also called  dilaton gravity) is coupled to the 
$\langle T_{ab}\rangle$ of quantized 2D massless and minimal
scalar fields by some sort 
of backreaction equations \cite{BaCR}. 
The physically more promising ones are those in which
the 2D dilaton gravity is the spherically symmetric 2D reduction of 4D 
Einstein's General Relativity \cite{PaPi},
where the dilaton $\phi$ is simply related to the 
radius of the two spheres $r$ by means of the simle relation
$r=e^{-\phi}$.  
The effective theory one is then considering is described by a 2D action of 
the form  
\be
S= S_{cl} +S_{P},
\label{azi}
\ee
where 
\be
S_{cl}=\frac{1}{2\pi }\int d^2 x \sqrt{-g^{(2)}}e^{-2\phi}[R^{(2)} + 
2(\nabla\phi)^2 +2 e^{2\phi}]
\label{eire}
\ee
 is the spherically symmetric reduction 
of 4D Einstein gravity 
%$S_E=\frac{1}{8\pi }\int d^4 x \sqrt{-g^{(4)}}R^{(4)}$, 
%\be
%S_m^{(2)}= -\frac{1}{4\pi }\int d^2 x \sqrt{-g^{(2)}}\sum_{i=1}^{N}(\nabla
%f_i)^2
%\label{azma}
%\ee
%is the scalar matter action 
and $S_P$ is the so called Polyakov action \cite{Polya}
and will be given in section 2.
This approach may be criticised since while the first term, $S_{cl}$, 
has a real 4D origin
the same can not be said for $S_P$. Coupling 4D spherically symmetric 
General Relativity to 2D quantum fields appears to be rather naive. 
In a more solid approach to the spherically symmetric case also the quantum 
fields should come from dimensional reduction of 4D. The idea is then 
to start 
from 4D minimally coupled scalar fields, perform the dimensional reduction 
under spherical symmetry and evaluate an effective 2D action for 
this kind matter to replace $S_P$ in 
 eq.  (\ref{azi}) \cite{MuWiZe}, \cite{effac}. 
This 2D effective action, which we will call  $S_{aind}$, is constructed
by functionally integrating the trace anomaly (see also \cite{anom},
\cite{KuLiVa}, \cite{NoOd}). 
The hope is to obtain in this way a more realistic picture of black 
hole evaporation. \\
Unfortunately, for the Schwarzschild space-time the $\langle T_{ab}\rangle$
so deduced is not even in qualitative agreement with the 4D
$\langle T_{\mu\nu}\rangle$ : it predicts a negative (!)
Hawking flux for an evaporating black hole \cite{MuWiZe}, 
\cite{KuLiVa}. 
This fact shades serious doubts
on the validity of this more ``sophisticated'' (as compard to eq.  (\ref{azi}))
2D approach. Puzzled with this problem, the authors of Ref. \cite{MuWiZe}
proposed to add Weyl invariant nonlocal terms to the above 
effective action $S_{aind}$. The resulting $\langle T_{ab}\rangle$ 
has the desired feature for a Schwarzschild black hole and 
 correctly reproduces the Hawking flux at infinity. However, for Minkowski
space-time this $\langle T_{ab}\rangle$ badly diverges and Minkowski 
is not a solution of the backreaction equations. \par \noindent
The situation appears rather frustrating. Trying to improve the simple 
Polyakov action to obtain a more accurate description of physical black holes
one gets unacceptable results. However, the physical motivations for these
improvements seem very reasonable. The possibility of implementing
 $S_{aind}$ also by means of Weyl invariant terms should neverthless 
lead to a stress tensor that in our opinion has to satisfy the following 
requirements:   
\par \noindent
i) conservation equations (in 4D);
\par \noindent
ii) vanishing in the vacuum Minkowski spacetime;
\par \noindent
iii) a  4D trace anomaly that, like in eq.  (\ref{tial}), does not depend 
on the state in which the expectation values are taken.
\par \noindent
In this paper we shall propose a $\langle T_{\mu\nu}\rangle$
derived in part by $S_{aind}$ that indeed satisfies the above 
requirements and is also in good qualitative agreement with the 4D
results.
\\
In order to make our analysis more clear, in sections 2 and 3 we will 
rederive with some detail all the known results about the Polyakov 
theory applied to the Schwarzschild black hole. We think that this 
part is necessary in order to understand better the rest of paper.
In section 4 we will then apply the two-dimensional techniques
just introduced to $S_{aind}$ and see how they lead to 
physically inconsistent results (such a systematic derivation
of these results is not present in the literature). In section 5
we show why we think that the effection action proposed in 
\cite{MuWiZe} does not improve much the situation.
In the last two paragraphs, 6 and 7, we will propose a possible 
solution to the problem based on the four dimensional interpretation.  
%%%%%%%%%%%%%%%%%%%%%%%%%%%%%%%%%%%%%%%%%%%%%%%%%%%%%%%%%%%%%%%%%%
\sect{Minimally coupled 2d fields}
%%%%%%%%%%%%%%%%%%%%%%%%%%%%%%%%%%%%%%%%%%%%%%%%%%%%%%%%%%%%%%%%%%
The action describing a conformally and minimally coupled scalar field 
$f$ in 2D is 
\be
S_m^{(2)}= -\frac{1}{4\pi }\int d^2 x \sqrt{-g^{(2)}}(\nabla
f)^2 \ ,
\label{uu}
\ee
leading to the field equation 
\be
\Box f =0 .
\label{aa}
\ee
Quantization is achieved by expanding the field operator $\hat f$ in
normal modes. Being every 2D metric locally conformally flat, one can 
introduce a coordinate system (not unique) in which the metric takes the form
\be
ds^2 =- e^{2\rho}dx^+dx^- .
\label{ac}
\ee
We shall call this system the $\{ x^{\pm} \}$ conformal frame. 
In this frame normalized positive frequency mode functions are of the form
\be
(4\pi w)^{-1/2}e^{-iwx^+}, \ \ \ (4\pi w)^{-1/2}e^{-iwx^-} .
\label{lo}
\ee
Expansion of $\hat f$ in the basis (\ref{lo}) selects a conformal state,
call it $|x^{\pm}\rangle$, in which the expectation value of the renormalized 
stress energy tensor operator for the scalar fields is  
\cite{davful}
\ba
\langle x^{\pm} |T_{\pm\pm}|x^{\pm}\rangle &=& - \frac{1}{12\pi}
(\partial_{\pm}\rho\partial_{\pm}\rho - \partial_{\pm}^2\rho),
\nonumber \\
\langle x^{\pm}|T_{+-}|x^{\pm}\rangle &=& 
-\frac{1}{12\pi} \partial_+\partial_-\rho \ .
\label{renst}
\ea
This stress tensor, as can be easily checked, is conserved, namely 
\be
\partial_{\mp}\langle x^{\pm}| T_{\pm\pm}|x^{\pm} \rangle
+\partial_{\pm}\langle x^{\pm}| T_{+-}|x^{\pm}\rangle  - \Gamma^{\pm}_{\pm\pm}
\langle x^{\pm}|T_{+-}|x^{\pm}\rangle =0
\label{vi}
\ee 
but it has, unlike its classical counterpart, developed a trace 
\be
\langle x^{\pm}|T|x^{\pm}\rangle = \frac{R^{(2)}}{24\pi}
\label{tutr}
\ee
where $T\equiv T^a_a$ and $R^{(2)}$ (hereafter $R$) is the Ricci scalar 
associated to the metric $g_{ab}^{(2)}$. 
This is the so called trace anomaly \cite{Duff}, which signals the 
breaking of conformal invariance at the quantum level. \\
The choice of the
normal modes (\ref{lo}) is by no means unique. One should equally
well had chosen another set of normal modes, obtained by solving the 
field equation (\ref{aa}) in another conformal frame, say 
$\{ \tilde x^{\pm} \}$ where $\tilde x^{\pm}=\tilde x^{\pm} (x^{\pm})$ , i.e.
\be
(4\pi \tilde w)^{-1/2} e^{-i\tilde w \tilde x^+}, \ \ \ 
(4\pi \tilde w)^{-1/2} e^{-i\tilde w \tilde x^-}.
\label{cari}
\ee
The expansion of the field operator $\hat f$ 
in terms of these new modes selects a 
conformal state $|\tilde x^{\pm}\rangle$. The expectation value of the 
energy momentum tensor (EMT) in this state is \cite{Balbinot}
\ba
\langle \tilde x^{\pm}| T_{\pm\pm}|\tilde x^{\pm} \rangle &=&
\langle x^{\pm}| T_{\pm\pm}| x^{\pm} \rangle +\Delta_{\pm}(x^{\pm}), 
\nonumber \\
\langle \tilde x^{\pm}| T_{+-}|\tilde x^{\pm} \rangle &=&
\langle x^{\pm}| T_{+-}| x^{\pm} \rangle .
\label{cacar}
\ea
Here 
\be
\Delta_+ (x^+)=\frac{1}{24\pi} \left( \frac{G''}{G}- \frac{1}{2}
\frac{G^{'2}}{G^2} \right)\ ,
\label{deu}
\ee
where 
\be
G(x^+)=\frac{dx^+}{d\tilde x^+}
\label{deg}
\ee
and a prime indicates derivation with respect to $x^+$. 
Similarly, 
\be
\Delta_- (x^-)=\frac{1}{24\pi} \left(\frac{F''}{F}- \frac{1}{2}
\frac{F^{'2}}{F^2} \right)
\label{ded}
\ee
and 
\be 
F(x^-)=\frac{dx^-}{d\tilde x^-} .
\label{def}
\ee
Note that in eqs. (\ref{cacar}) the components of the stress tensor are still 
expressed, as in (\ref{renst}), in the $\{ x^{\pm} \}$ frame, but the 
expectation value is taken in the $|\tilde x^{\pm}\rangle$ state.
From eqs. (\ref{deu}), (\ref{ded}) we can see that the $\Delta_{\pm}$
is proportional to the Schwarzian derivative between $x^{\pm}$ and 
$\tilde x^{\pm}$. From the physical point of view, $\Delta_{\pm}$ give
the expectation values of $T_{\pm\pm}$ in the state $|\tilde x_{\pm}\rangle$
normal ordered with respect to  $| x_{\pm}\rangle$. \\
We see from eqs. (\ref{cacar}) that $\Delta_{\pm}$ 
represents conserved massless (i.e. trace free) radiation and is the 
only difference in the expectation values of $\langle T_{ab}\rangle$ 
in two distinct conformal states, being the trace anomaly state independent 
(see eq. (\ref{tutr})). This difference is nonlocal in the sense that it 
does not depend on the local geometry, but rather on the global definition 
(through the normal modes) of the states. 
Therefore $\Delta_{\pm}$ represent a nonlocal contribution to  
 $\langle T_{ab}\rangle$ that depends on the state in which the expectation 
values are taken. 
\par \noindent
The expectation values of the EMT (eqs. (\ref{renst}))
can also be easily obtained by integrating the conservation equations
$\nabla _a\langle T^a_b\rangle =0$ once the trace anomaly eq. (\ref{tutr})
is given \cite{Davies}. 
In the generic conformal frame of eq. (\ref{ac}) the only non vanishing 
Christoffel symbols are 
\be
\Gamma^{\pm}_{\pm\pm} = 2\partial_{\pm}\rho .
\label{chsi}
\ee
Inserting  this and the second of eqs. (\ref{renst}) in the conservation 
equations (\ref{vi}), straightforward integration leads to 
\be
\langle T_{\pm\pm}\rangle = - \frac{1}{12\pi} (\partial_{\pm}\rho
\partial_{\pm}\rho - \partial_{\pm}^2\rho - t_{\pm}(x^{\pm}) ),
\label{rigi}
\ee
where $ t_{\pm}(x^{\pm})$ are two arbitrary integration functions of their
respective arguments. They signal the nonlocal character of 
$\langle T_{\pm\pm}\rangle$ because of its state dependence. In view of the 
preceeding discussion, $ t_{\pm}(x^{\pm})$ are related to the 
Schwarzian derivatives $\Delta_{\pm}(x^{\pm})$ of eqs. 
(\ref{cacar}). 
Note that, as the trace anomaly does not depend on the quantum state, the 
$ t_{\pm}(x^{\pm})$ are necessary, in eq. (\ref{rigi}), to specify in 
which quantum state the expectation values are taken.
Another way of seeing the appearance of these terms is to consider that 
under the transformation $x^{\pm}\to \tilde x^{\pm}$, which is at the same 
time a conformal and a coordinate transformation, $\langle T_{ab}\rangle$ 
does not transform as a tensor. Because of the breaking of the conformal
invariance at the quantum level, the transformation of $\langle T_{ab}\rangle$
involves an anomalous contribution, namely the Schwarzian derivative. 
\par \noindent
An elegant way of recovering the previous results is to functionally
integrate the trace anomaly eq. (\ref{tutr}) obtaining Polyakov's 
nonlocale effective action \cite{Polya} 
\be
 S_P=-\frac{1}{96\pi } \int d^2 x \sqrt{-g} R \frac{1}{\Box} R \ , 
\label{viii}
\ee
where $\Box$ is the covariant Dalambertian. Varying $S_P$ with respect to 
$g^{ab}$ gives
\ba
\langle T_{ab}\rangle &=& -\frac{1}{96\pi } \Big\{
-2\nabla_{a}\nabla_{b}
( \frac{1}{\Box} R )+\nabla_{a} (\frac{1}{\Box} R )\nabla_b (\frac{1}{\Box R})  
\nonumber \\
&&+g_{ab} \left[ 2R - \frac{1}{2 }\nabla^{c}
 ( \frac{1}{\Box} R )\nabla_{c} ( \frac{1}{\Box} R ) \right]\Big\} \ .
\label{ix}
\ea
Choosing now a conformal frame eq. (\ref{ac}), where 
$\frac{1}{\Box}R=-2\rho$, we recover the previous expressions for 
$\langle T_{ab}\rangle$, namely eqs. 
(\ref{renst}). The functions $t_{\pm}(x^{\pm})$ can always be added to 
$\langle T_{ab}\rangle$ because they are compatible with the conservation
equations. Furthemore, as before being the trace anomaly state independent,
$S_P$ is the same for every quantum state. 
So the inclusion of $t_{\pm}$ is necessary to specify the state.   
Within this respect, one should also note that the trace anomaly 
determines only the Weyl non invariant part of the effective action, namely
$S_P$. The complete effective action could in principle contain also 
Weyl invariant nonlocal terms. These, however, do not contribute 
to the trace $\langle T\rangle$ and by requiring the conservation 
equations one concludes that their contribution to $\langle T_{\pm\pm}\rangle$
should be of the form $t_{\pm}(x^{\pm})$. 
\par \noindent
As a final remark, one should remind that the renormalization procedure 
which leads to $S_P$ in principle  also allows for the presence of a two 
dimensional cosmological constant 
term \cite{Polya}. The importance of such a term
%for non asymptotically flat space-times 
will be considered in section 6.
%%%%%%%%%%%%%%%%%%%%%%%%%%%%%%%%%%%%%%%%%%%%%%%%%%%%%%%%%%%%%%%%%
\sect{The 2D Schwarzschild black hole}
%%%%%%%%%%%%%%%%%%%%%%%%%%%%%%%%%%%%%%%%%%%%%%%%%%%%%%%%%%%%%%
We now apply the results of the previous section to the 2D Schwarzschild 
black hole. In the Eddington-Finkelstein null frame $\{ u,v \} $ the 
metric reads
\be
ds^2 =- (1-\frac{2M}{r})dudv \ ,
\label{edfi}
\ee
where
\be
v=t+ r_* , \ \ \ u=t-r_*
\label{nuco}
\ee
and 
\be
r_*= \int \frac{dr}{1-2M/r} = r+ 2M \ln |\frac{r}{2M}-1| \ .
\label{cota}
\ee
$M$ represents the mass of the black hole. 
Expansion of the field operator $\hat f$ in the modes
\be
(4\pi w)^{-1/2} e^{-iwv}, \ \ \ (4\pi w)^{-1/2}e^{-iwu}
\label{moef}
\ee
defines a conformal state known as the Boulware vacuum $|B\rangle$. 
Application of eqs. (\ref{renst}) gives \cite{Fulling}
\ba
\langle B|T_{uu}|B\rangle &=& \langle B|T_{vv}|B\rangle = 
\frac{1}{24\pi} [ - \frac{M}{r^3} +\frac{3}{2} \frac{M^2}{r^4} ]   ,
\nonumber \\
\langle B|T_{uv}|B\rangle &=& -\frac{1}{24\pi} (1- \frac{2M}{r})\frac{M}{r^3}.
\label{budu}
\ea
As one immediately sees, the modes in (\ref{moef}) reduce at infinity to the 
usual Minkowski ingoing and outgoing plane waves and there 
$\langle B| T_{ab} |B\rangle =0$. So the state $|B\rangle$ reproduces at
infinity the familiar notion of an empty vacuum state as inferred from 
Minkowski field theory. One can think of this feature as the reason for 
selecting  $|B\rangle$ among the various candidates for a reasonable 
vacuum state of the theory. \\
If the behaviour of $|B\rangle$ at infinity seems quite reasonable, the same 
cannot be said for the horizon $r=2M$. One expects in fact that, if 
these regions belong to the physical space-time manifold, 
$\langle T_{ab}\rangle$  should be finite there with respect to a local 
orthonormal frame. It can be shown that $\langle T_{ab}\rangle$ is regular
on the future horizon if as $r\to 2M$ \cite{ChFu}
\ba
| \langle T_{vv} \rangle | &<& \infty , \nonumber \\
(r-2M)^{-1} | \langle T_{uv} \rangle | &<& \infty , \nonumber \\
(r-2M)^{-2} | \langle T_{uu} \rangle | &<& \infty \ .
\label{core}
\ea
The regularity on the past horizon is expressed by similar inequalities 
with $u$ and $v$ interchanged. 
It is now clear that $\langle B|T_{ab} |B \rangle$ is not regular both on the 
future and the past horizons. This behaviour is connected to the fact that 
the state $|B\rangle$ is defined in terms of the $(u,v)$ modes in 
(\ref{moef}) which oscillate infinitely on the horizon. 
Physically, the state $|B\rangle$ is supposed to describe the vacuum 
polarization of the space-time exterior to a static massive body 
whose radius is bigger than $2M$. 
\par \noindent
A coordinate system regular on the horizons is the Kruskal $\{ U,V \} $
one defined in terms of the $\{ u,v \} $ frame as (for $r>2M$) 
\be
U=-4Me^{-u/4M}, \ \ \ V=4Me^{v/4M} \ .
\label{xiii}
\ee
Expansion of the field operator $\hat f$ in the Kruskal modes
\be
(4\pi \tilde w)^{-1/2}e^{-i\tilde w V}, \ \ \ 
(4\pi \tilde w)^{-1/2}e^{-i\tilde w U}
\label{numo}
\ee
defines the Hartle-Hawking state $|H\rangle$. 
Evaluating the Schwarzian derivatives between $U(V)$ and $u(v)$, from eqs. 
(\ref{cacar})-(\ref{def}) we obtain 
\ba
\langle H|T_{uu}|H\rangle = \langle H|T_{vv}|H\rangle &=&
\frac{1}{768\pi M^2} (1- \frac{2M}{r})^2 [1+\frac{4M}{r} + \frac{12M^2}{r^2} ],
\nonumber \\
\langle H|T_{uv}|H\rangle &=&- \frac{1}{24\pi} (1- \frac{2M}{r})\frac{M}{r^3}.
\label{emku}
\ea
This state leads therefore to expectation values regular on both future 
and past horizons. 
The Kruskal modes, however, do not reduce asymptotically to standard Minkowski 
plane-waves. As a consequence, $\langle H|T_{ab}|H\rangle$ does not
vanish at infinity. $|H\rangle$ is a thermal state at the Hawking temperature
\be
T_H=\frac{1}{8\pi M}
\label{tani}
\ee
and describes the thermal equilibrium of a black 
hole enclosed in a box with its radiation. 
\par \noindent
The last example we shall present deals, unlike the previous ones, with a 
dynamical situation, namely the formation of a black hole by gravitational
collapse. It will be of fundamental relevance for the subsequent discussion.
Let us consider, to limit the mathematical complexity, the simple case 
where the black hole is formed by the collapse of a shock-wave at $v=v_0$
\cite{Hisc} (for the timelike case see for instance \cite{fdu}).
In the ``in'' region $v<v_0$ the space-time is flat 
\be
ds_{in}^2=-du_{in}dv_{in} ,
\label{xvii}
\ee
where $u_{in}$ and $v_{in}$ are the usual retarded and advanced Minkowski 
null coordinates 
\be
u_{in} = t_{in} - r_{in}, \ \ \ v_{in}=t+r_{in}.
\label{miar}
\ee
For $v>v_0$ the ``out'' geometry describes a black hole of mass $M$
\be
ds^2_{out}= - (1-2M/r)dudv \ .
\label{oubl}
\ee
Matching the two geometries at $v=v_0$ we have 
\ba 
v &=& v_{in}, \nonumber \\
u &=& u_{in} -4M \ln \left(\frac{v_0 -u_{in} -4M}{4M}\right).
\label{maeq}
\ea
We choose the quantum state to correspond to Minkowski vacuum on past null 
infinity. Call this state $|in\rangle$. Therefore 
$\langle in| T_{ab}|in\rangle  =0 $ for $v<v_0$. 
\par \noindent
The evaluation of the expectation values for $v>v_0$ requires the Schwarzian 
derivative between $u$ and $u_{in}$. The net result is, for $v>v_0$,
\ba
\langle in |T_{uu}|in\rangle &=& \frac{1}{24\pi}
\left[ - \frac{M}{r^3} + 
\frac{3}{2}\frac{M^2}{r^4} - \frac{8M}{(u_{in} -v_0)^3}
-\frac{24M^2}{(u_{in}-v_0)^4} \right] \ , \nonumber \\
\langle in | T_{vv}| in\rangle &=& \frac{1}{24\pi} 
\left[ - \frac{M}{r^3} + \frac{3}{2}\frac{M^2}{r^4} \right] =  
\langle B| T_{vv} | B\rangle \ , \nonumber \\
\langle in | T_{uv}| in\rangle &=& -\frac{1}{24\pi} (1-\frac{2M}{r})
\frac{M}{r^3} = \langle B| T_{uv} | B\rangle \ .
\label{esin}
\ea
In the limit $u_{in}\to v_0 - 4M$ (i.e. the shell is close to crossing 
the horizon) at infinity we find a net flux 
\be
\langle T_{uu}\rangle \to \frac{1}{768\pi M^2} 
\label{fleva}
\ee
representing the Hawking flux of evaporation at the correct Hawking temperature 
$T_H$. In the above limit ($u_{in}\to v_0 - 4M$) all time dependence 
disappears  
\be
\langle in | T_{uu} |in \rangle = \frac{1}{768\pi M^2} (1-\frac{2M}{r})^2
\left[ 1+ \frac{4M}{r} + \frac{12M^2}{r^2} \right] 
\label{noti}
\ee
and the state $| in\rangle $ becomes what is called the Unruh state 
$|U\rangle $,
which is obtained by expanding the field operator $\hat f $ 
in modes obtained by using the coordinate $U$ for the outgoing modes 
and the coordinate $v$ for the ingoing ones. 
Note that $\langle U| T_{ab}|U\rangle $ is regular on the future horizon
(see eqs. \ref{core}). The singularity on the past horizon is 
completely spurious since for the black hole formed by gravitational 
collapse there is no past horizon.
\par \noindent
These three examples were given not only as an application of the formalism,
but rather to show how for the Schwarzschild space-time the two-dimensional
$\langle T_{ab}\rangle$ given by the Polyakov action reproduces 
the basic qualitative features of the 4D $\langle T_{\mu\nu}\rangle$.
This qualitative agreement of the simple two dimensional calculations 
with the more complicated four dimensional ones is quite amazing.
It has stimulated investigations of the backreaction problem, 
namely the evolution of the black hole as it emits Hawking radiation, 
by means of two dimensional models where the classical 
dilaton-gravity action is improved by adding the Polyakov term $S_P$.
This gives an effective action which describes the effect of the quantized 
matter  on the geometry, i.e. the backreaction. The same problem can not
even be attacked in the physical 4D context. 
\par \noindent
The coupling of the minimal 2D massless scalar fields to the spherically 
symmetric reduced Einstein-Hilbert action (or similar) to evaluate the 
backreaction might sound too naive. 
One can cogently argue that in a ``realistic'' 2D matter-gravity theory
also the matter sector should derive, through dimensional reduction, from 
a consistent 4D theory. 
However, before embarking in backreaction calculations one should assure 
that these more ``sophisticated'' 2D models produce a 
$\langle T_{ab}\rangle$ which for the Schwarzschild space-time 
is at least in qualitative agreement with the 4D $\langle T_{\mu\nu}\rangle$
as it was for the naive Polyakov theory. 
Otherwise these ``sophisticated'' models suffer from physical 
inconsistency. 
%%%%%%%%%%%%%%%%%%%%%%%%%%%%%%%%%%%%%%%%%%%%%%%%%%%%%%%%%%%%%%%%
\sect{Minimally coupled 4D fields}
%%%%%%%%%%%%%%%%%%%%%%%%%%%%%%%%%%%%%%%%%%%%%%%%%%%%%%%%%%%%%%%%
As we have seen, while the gravitational part of the action has a 
four dimensional origin, the matter sector is two dimensional. 
It seems therefore natural, in the search for a more physical model, 
to require that also the matter fields 
should be defined ab initio in 4D \cite{MuWiZe}, \cite{effac}, 
\cite{anom}, 
\cite{KuLiVa}, \cite{NoOd}.
Restricting the attention to minimally coupled 4D scalar fields, one has 
that the corresponding 4D action reads 
\be
S_M^{(4)}= - \frac{1}{(4\pi)^2} \int d^4 x \sqrt{-g^{(4)}}
(\nabla f)^2 \ .
\label{maqu}
\ee
Under the assumption of spherical symmetry the 4D metric can be written as 
\be
ds^2 = g_{ab}dx^a dx^b + e^{-2\phi}d\Omega^2\ ,
\label{sfer}
\ee
where $g_{ab}(x^a)$, $a,b=1,2$, is the two-dimensional metric and $d\Omega^2$
the line element of the unit two-sphere. 
Performing the dimensional reduction in eq. (\ref{maqu}) and using 
eq.  (\ref{sfer}) we arrive at a 2D action for our scalar fields 
\be
S_M^{(2)}= -\frac{1}{4\pi }\int d^2 x \sqrt{-g^{(2)}}
e^{-2\phi}(\nabla
f)^2 \ ,
\label{v}
\ee 
leading to the field equation 
\be
\nabla^{\alpha}(e^{-2\phi}\nabla_{\alpha}f)=0 \ .
\label{theq}
\ee
Comparing the action  (\ref{v}) to (\ref{uu}), we see that the scalar 
fields, from a 2D point of view, are still conformal but there is now a 
coupling between them and the dilaton $\phi$. This makes the trace anomaly 
to differ from eq.  (\ref{tutr}) by extra $\phi$ terms 
\cite{MuWiZe}, \cite{effac},
\cite{anom}, \cite{KuLiVa}, \cite{NoOd}
\be
\langle T\rangle =\frac{1}{24\pi}\left[R-6(\nabla\phi)^2+6 \Box \phi
\right]\ ,
\label{xxiii}
\ee
which is still state independent. 
The coefficient of the last term in eq. (\ref{xxiii}) is not
unambiguously given in the literature, depending on the functional
measure used for the scalar fields, i.e. genuine 2D versus spherically
symmetric reduced $4D$. The last choice is the one which leads to our 
eq. (\ref{xxiii}). 
%Needless to say that any other choice would 
%lead to a nonvanishing  $\langle T\rangle$ in Minkowski space.   
In a generic conformal frame $\{ x^{\pm} \}$ we have
\be
\langle T_{+-}\rangle = -\frac{1}{12\pi}\left(\partial_+\partial_-\rho
+3\partial_+\phi\partial_-\phi-3\partial_+\partial_-\phi\right).
\label{xxiv}
\ee
\par \noindent
The problem we have to face now is to construct the other components of 
$\langle T_{ab}\rangle$ for this improved theory. 
Following the analysis of the previous section, one could integrate 
the 2D conservation equations $\nabla_a \langle T^a_b\rangle =0$, obtaining 
(this is the approach of \cite{KuLiVa} )
\ba
\langle T_{\pm\pm}\rangle &=& -\frac{1}{12\pi}\left(\partial_{\pm}\rho 
\partial_{\pm}\rho -\partial_{\pm}^2\rho - t_{\pm}\right) 
-\frac{1}{4\pi}\left[\frac{1}{\partial_{\pm}}
\left( 2\partial_{\pm}\rho\partial_+\phi
\partial_{-}\phi\right) -\frac{\partial_{\pm}}{\partial_{\mp}}
(\partial_-\phi\partial_+\phi) \right]\nonumber \\
&&+\frac{1}{4\pi}\left[ \frac{1}{\partial_{\mp}}
\left(2\partial_{\pm}\rho\partial_+\partial_-\phi\right)
-\partial_{\pm}^2\phi \right] \ ,
\label{xxv}
\ea
where we have used the shorthand notation
\be
\frac{1}{\partial_{\pm}}= \int dx^{\pm}\ .
\label{shno}
\ee
The functions $t_{\pm}(x^{\pm})$ in eqs. (\ref{xxv}) are arbitrary integration
functions. Comparing eqs. (\ref{xxv}) with eqs. (\ref{renst}) 
we see the appearance of dilaton dependent terms.
\par \noindent 
The other approach that we can follow, again as in the previous section, 
is to functionally integrate the trace anomaly eq. (\ref{xxiii}) to 
obtain the 2D effective action (anomaly induced effective action)
 \cite{MuWiZe}, \cite{effac}, \cite{NoOd} 
\be
S_{aind}=-\frac{1}{2\pi}\int d^2 x \sqrt{-g}\left[\frac{1}{48} R
\frac{1}{\Box}R
-\frac{1}{4}(\nabla\phi)^2 \frac{1}{\Box} R + \frac{1}{4}\phi R\right].
\label{xxvii}
\ee
The first term in this action is $S_P$ of eq. (\ref{viii}) , the Polyakov 
action. However, a new nonlocal term  has now appeared in the effective 
action, the second one. The last term, on the contrary, is local.
By varying $S_{aind}$ with respect to the 2D metric $g_{ab}$ we find
\ba
\langle T_{ab}\rangle &=& \langle T_{ab}^P\rangle +\frac{1}{8\pi}
\Big\{ -\frac{g_{ab}}{2}\left[ (\nabla\phi)^2 (\frac{1}{\Box} R)+ 
\nabla^{c}\left( \frac{1}{\Box} (\nabla\phi)^2
\right) \nabla_{c}(\frac{1}{\Box} R)
-2(\nabla\phi)^2 \right] \nonumber \\
&& +\partial_{a}\phi\partial_{b}\phi (\frac{1}{\Box} R) 
 + \frac{1}{2}\left[ \nabla_{a}
\left(\frac{1}{\Box} (\nabla\phi)^2 \right)\nabla_{b}(\frac{1}{\Box}R)
+  \nabla_{b}
\left(\frac{1}{\Box} (\nabla\phi)^2
\right)\nabla_{a}(\frac{1}{\Box}R)\right]  
\nonumber \\
&&-\nabla_{a}\nabla_{b}\left(\frac{1}{\Box}(\nabla\phi)^2\right)\Big\}
-\frac{1}{8\pi}   
\left(g_{ab}\Box\phi -\nabla_{a}\nabla_{b}\phi\right) ,
\label{xxviii}
\ea
where $\langle T_{ab}^P\rangle$ was given eqs. (\ref{ix}) and comes
from the Polyakov term in $S_{aind}$. In the conformal frame 
$\{ x^{\pm} \}$ eqs. (\ref{xxviii}) read
 (see also
\cite{NoOd})
\be 
\langle T_{\pm\pm}\rangle = \langle T_{\pm\pm}^P \rangle +
\frac{1}{2\pi} \left[ \rho\partial_{\pm}\phi\partial_{\pm}\phi  
+\frac{1}{2}
\frac{\partial_{\pm}}{\partial_{\mp}}(\partial_+\phi\partial_-\phi)\right]
-\frac{1}{4\pi}
\left(-2\partial_{\pm}\rho\partial_{\pm}\phi+\partial_{\pm}^2\phi\right)
\label{xxix}
\ee
and the trace is obviously eq. (\ref{xxiii}).
Surprisingly, the two expressions of $\langle T_{ab}\rangle$ given, 
namely eqs. (\ref{xxv}) obtained by integrating the 2D conservation equations 
$\nabla_a \langle T^a_b \rangle =0$ and eqs. (\ref{xxix}) obtained 
by functional differentiation of $S_{aind}$, do not coincide (whatever the 
functions $t_{\pm}$ might be).
We shall see in section 6 that the procedure followed to get eqs. (\ref{xxv})
is not justified. Therefore we shall discuss, here, only the 
$\langle T_{ab}\rangle$ given by eqs. (\ref{xxiv})  and (\ref{xxix}).
\par \noindent
Before starting the calculation of $\langle T_{ab}\rangle$ for the 
Schwarzschild black hole, one has to implement eq. (\ref{xxix}) by a state 
dependent term which selects the state in which the expectation values
are taken. 
Naively, one could just add a term $t_{\pm}(x^{\pm})$ as in the previous 
section, since it is compatible with the 2D conservation equations 
satisfied by the Polyakov term $\nabla_a \langle T^{Pa}_b\rangle =0$. 
However more care is required. Extra terms in $\langle T_{\pm\pm}\rangle$
arise as a consequence of the ``anomalous'' transformation of 
$\langle T_{ab}\rangle $ under the transformation $x^{\pm}\to \tilde x^{\pm}$.
\\
Let us identify the previous expression eqs. (\ref{xxix}) as expectation values
of $T_{ab}$ in the state $|x^{\pm} \rangle$, i.e. 
$\langle x^{\pm}| T_{ab}|x^{\pm}\rangle$. Consider now, as we did before, the 
$|\tilde x^{\pm}\rangle$ state. We already know how the first term in eq. 
 (\ref{xxix}), the Polyakov one, transforms. 
It is also easy to verify that the terms obtained by 
variation of $\phi R$ in $S_{aind}$ (like the trace) is 
state independent. We come now to the term in square bracket in 
eq.  (\ref{xxix}), the second one; call it $T_{\pm\pm}^{(2)}$. We find
\be
\langle \tilde x^{\pm}|T_{--}^{(2)}|\tilde x^{\pm}\rangle =
\langle x^{\pm}|T_{--}^{(2)} |x^{\pm}\rangle + 
\frac{1}{4\pi} \left[ \partial_{-}\phi \partial_-\phi \ln ( FG) +
\frac{F'}{F} \int dx^+ \partial_+\phi\partial_-\phi \right]
\label{strao}
\ee
and similarly for $T_{++}^{(2)}$ by interchanging $-$ with $+$ 
and $F$ with $G$. $F$ and $G$ are defined as in eqs. (\ref{deg}) and
(\ref{def}). 
Summing up we have that for the theory described by $S_{aind}$ 
\ba
\langle \tilde x^{\pm} | T_{++}|\tilde x^{\pm}\rangle &=& 
\langle x^{\pm} | T_{++}| x^{\pm}\rangle + \frac{1}{24\pi}
\left( \frac{G''}{G} - \frac{1}{2}\frac{G^{'2}}{G^2} \right)
\nonumber \\
+ && \frac{1}{4\pi}
\left[ \partial_{+}\phi \partial_+\phi \ln ( FG) +
\frac{G'}{G} \int dx^- \partial_+\phi\partial_-\phi \right]  ,
\label{gul}
\ea
\ba
\langle \tilde x^{\pm} | T_{--}|\tilde x^{\pm}\rangle &=& 
\langle x^{\pm} | T_{--}| x^{\pm}\rangle + \frac{1}{24\pi}
\left( \frac{F''}{F} - \frac{1}{2}\frac{F^{'2}}{F^2} \right)
\nonumber \\
+ && \frac{1}{4\pi}
\left[ \partial_{-}\phi \partial_-\phi \ln ( FG) +
\frac{F'}{F} \int dx^+ \partial_+\phi\partial_-\phi \right],
\label{gur}
\ea
\be
\langle \tilde x^{\pm}| T_{-+} |\tilde x^{\pm}\rangle =
\langle  x^{\pm}| T_{-+} |  x^{\pm}\rangle \ .
\label{gua}
\ee
So going from one conformal state to another $\langle T_{ab}\rangle$
does not only acquire a term proportional to the Schwarzian derivative, 
but also the last two terms in eqs. (\ref{gul}) and (\ref{gur}). 
These do not represent, unlike $t_{\pm}$, massless 2D radiation 
and are much more complicated in this more ``sophisticated'' 2D model.
Being these new state dependent terms nonlocal, there is a serious danger 
that they destroy the nice qualitative agreement in a Schwarzschild 
background between the prediction of the Polyakov EMT 
$\langle T_{ab}^P\rangle$ and the 4D $\langle T_{\mu\nu}\rangle$.
In order to see if this is the case, we now calculate, using 
eqs. (\ref{gul}), (\ref{gur}) and (\ref{gua}), the $\langle T_{ab}\rangle$
for the three states ($|B\rangle$, $|H\rangle$ and $|U\rangle$) 
defined on the Schwarzschild space-time in section 3
and compare the result we obtain to the 4D $\langle T_{\mu\nu}\rangle$
described in the introduction. 
\par \noindent
For the Boulware state $|B\rangle$ we have $\tilde x^{\pm} = x^{\pm}=
(u,v)$ where $(u,v)$ are Eddington-Finkelstein coordinates 
( see eqs. (\ref{nuco}) ). Eqs. (\ref{gul}), (\ref{gur}) and (\ref{gua})
give
\ba
\langle B|T_{uu}|B\rangle &=& \langle B|T_{vv}|B\rangle =
\frac{1}{24\pi}\left[ -\frac{M}{r^3} + \frac{3}{2} \frac{M^2}{r^4} 
\right] + \frac{1}{16\pi} (1-\frac{2M}{r})^2 \frac{1}{r^2} 
\ln (1-\frac{2M}{r}), \nonumber \\
\langle B|T_{uv}|B\rangle &=& -\frac{1}{24\pi} (1-\frac{2M}{r})\frac{M}{r^3}
+\frac{1}{8\pi} (1-\frac{2M}{r}) \frac{M}{r^3} \ .
\label{neub}
\ea
Note first that for $M=0$ $|B\rangle$ becomes the usual Minkowski 
vacuum $|M\rangle$ and eqs. (\ref{neub}) tell us that 
\be
\langle M|T_{ab}|M\rangle =0\ .
\label{neum}
\ee
This result, as we shall see in the next section, is not so trivial as it 
appears. It implies that Minkowski space-time is a consistent solution
of the semiclassical field equations. 
Needless to say that any other choice in the coefficient of the $\Box \phi$ 
term in the trace anomaly
would 
lead to a nonvanishing  $\langle T_{ab}\rangle$ in Minkowski space.   
Coming back to the 
Schwarzschild case ($M \neq 0$) we see that eqs. (\ref{neub}) contain,
in addition to the terms obtained by solely $S_P$, a term proportional
to $\frac{f^2}{r^2}\ln f$, where $f=1-\frac{2M}{r}$.
This gives, in Kruskal coordinates, a ``weak'' logarithmic divergence on the
horizon (see eqs. (\ref{core}) ). This divergence is however subleading 
when compared to the ``strong'' divergence $\sim \frac{1}{V^2}$ or 
$\sim \frac{1}{U^2}$ coming from $\langle B|T_{ab}^P|B\rangle$. 
Therefore, the physical features of the state $|B\rangle$ remain unaltered;
the ``sophisticated'' $S_{aind}$ introduces just extra vacuum polarization 
terms in the stress tensor in addition to those obtained by $S_P$. 
$|B\rangle$ can reasonably describe even in this theory the vacuum 
polarization of the space-time outside a static star. 
The qualitative agreement between $\langle B | T_{ab} | B\rangle$ 
and the 4D $\langle B|T_{\mu\nu}|B\rangle$ 
is still satisfactory.
\par \noindent
Let us now consider the state $|H\rangle$ obtained by choosing 
$\tilde x^{\pm}= (U, V)$ ( Kruskal coordinates, see eqs. (\ref{xiii}) )
and $x^{\pm}= (u,v)$. Here, as we shall see, things are not so ``nice'' as 
before. In this state we get, from eqs. (\ref{gul})-(\ref{gua}),
\ba
\langle H|T_{uu}|H\rangle &=& \langle H|T_{vv}|H\rangle = 
\frac{1}{768\pi M^2} (1-\frac{2M}{r})^2 (1+ \frac{4M}{r} +
\frac{12M^2}{r^2} ) \nonumber \\
&&+ \frac{1}{16\pi} \left[ (1-\frac{2M}{r})^2 \frac{1}{r^2} (-\ln r -
\frac{r}{2M}) - \frac{1}{2M} (-\frac{1}{r} + \frac{M}{r^2} + \frac{1}{4M})
\right], \nonumber \\
\langle H|T_{uv}|H\rangle &=& \langle B|T_{uv}|B\rangle \ .
\label{neuk}
\ea
Inspection of eqs. (\ref{neuk}) reveals that the $uu$ and 
$vv$ components of $\langle H|T_{ab}|H\rangle$    
vanish like $(r-2M)^2$ on the horizon. Therefore  
$\langle H|T_{ab}|H\rangle$  is regular on both the future and past 
horizons $r=2M$, as expected. 
However its behaviour as $r\to \infty$ is quite surprising   
\be
\langle H| T_{uu}|H\rangle = \langle H| T_{vv}|H\rangle 
\to \frac{1}{768\pi M^2} (1-6)\ ,
\label{rsup}
\ee
where the first term on the r.h.s. comes from $\langle T_{ab}^P\rangle$,
whereas the unexpected negative contribution (the $-6$) comes from the 
second nonlocal term in $S_{aind}$ (see eq. (\ref{xxvii})) \cite{MuWiZe}. 
This result looks rather unphysical, since it would suggest that the black 
hole is in thermal equilibrium with a thermal bath of negative energy.
Some clarifications are necessary to understand the validity of eqs. 
(\ref{neuk}) and their asymptotic limit. The lower bound $r_0$ of 
integration in the $r$ integral present in eqs. (\ref{gul}), 
(\ref{gur}) was taken to be $r_0=2M$. Any other choice (some of them
might eliminate the negativity of the net flux in the asymptotic limit eq.
(\ref{rsup})) leads to a $\langle H|T_{ab}|H\rangle$ singular on the 
horizon. For the state $|B\rangle$ the stress tensor does not depend 
on the choice of $r_0$. Furthemore, as said before, the coefficient in front
of the $\Box \phi$ term in the trace anomaly in eq. (\ref{xxiii}) has been 
source of debate in the literature. For the problem at hand we stress that
the $\Box\phi$ term affects only the local part of $\langle T_{ab}\rangle$
giving no extra contribution to the Hawking radiation and, therefore, has
nothing to do with 
the puzzling result we have in eq. (\ref{rsup}). 
So we have firm evidence that the 2D stress tensor $\langle T_{ab}\rangle$
constructed from $S_{aind}$ in the $|H\rangle$ state is in strong 
qualitative disagreement with the well established result of the 
4D $\langle H|T_{\mu\nu}|H\rangle$ which, we remind, describes a black 
hole in thermal equilibrium with a positive energy bath of radiation at the 
temperature $T_H$. \\
One can indeed find an equilibrium state $|\tilde x^{\pm}\rangle$ 
which is regular on the horizons and unlike $|H\rangle$ has a 
positive flux at infinity \cite{BuRaMi}. 
Mathematically this is done by fine tuning
two constants. One is the lower bound $r_0$ in the $r$ integration. 
The second, say $\alpha$, is related to the definition of the 
$\{ \tilde x^{\pm} \}$ frame 
\be
\tilde x^+ = \alpha e^{ v/ \alpha }, \ \ 
\tilde x^- = -\alpha e^{-u / \alpha }\ .
\label{nonk}
\ee
The choice of exponential relation is imposed by the need of having a 
constant Schwarzian derivative as required for equilibrium. 
The outgoing flux can then be parametrized by a third constant $\beta$ 
(which depends on $r_0$ and $\alpha$) as (in the limit $r\to \infty$)
\be
\langle \tilde x^{\pm}| T_{uu}|\tilde x^{\pm} \rangle =
\langle \tilde x^{\pm}| T_{vv}|\tilde x^{\pm} \rangle 
\sim   \frac{\beta }{768\pi M^2} \ .
\label{befl}
\ee
According to the 
previous calculations,
 no regular solution for $\beta=1$ exists. 
For $\beta\neq 1$ and positive one can find $\alpha$ ($\neq 4M$) and 
$r_0$ ($\neq 2M$) which allows regularity of $\langle \tilde x^{\pm}|
T_{ab} |\tilde x^{\pm}\rangle$ on the horizons. 
Needless to say that the state so constructed has nothing to do with $|H\rangle$ 
and its physical significance, if it exists, is completely obscure.
\par \noindent
Complete disagreement between the prediction of $S_{aind}$ and the real 4D 
theory emerges also when considering our last example: the collapsing shell.
Performing the calculation along the lines of the previous section 
we have 
\be
\langle in|T_{ab}|in\rangle =0
\label{curo}
\ee
for $v<v_0$. When, instead, $v>v_0$ 
\ba
\langle in|T_{uu}| in \rangle  &=&
\frac{1}{12\pi}\left(\frac{ff^{''}}{8}-\frac{f^{'2}}{16}
- \frac{3}{4}\frac{M^2}{r^4(u,v_0)} +
\frac{M}{2 r^3(u,v_0)}\right)+
\frac{1}{16\pi}\frac{f^2}{r^2}\Big[ \ln\frac{f}{f(u,v_0)}
\nonumber \\
&&-\frac{f^2(u,v_0)}{r^2(u,v_0)} +
\frac{2M}{r^2(u,v_0)}\left( -\frac{1}{r}+\frac{M}{r^2} +\frac{1}{
r(u,v_0)} -\frac{M}{r^2(u,v_0)} \right) \Big] ,  
\nonumber \\
\langle in| T_{vv}| in \rangle  &=&
\langle B| T_{vv}| B \rangle \ , \nonumber \\
\langle in| T_{uv}| in \rangle  &=&
\langle B| T_{uv}| B \rangle \ , 
\label{xxxiii}
\ea 
where  $f=1-2M/r$ and $r(u,v_0)=(v_0-u_{in})/2$.
In the first of eqs. (\ref{xxxiii}) the lower bound in the $v$ integration 
has been taken as $v_0$, the position of the shell, since this appears
as the more natural choice. The stress tensor is regular on the future 
horizon. As the shell approaches the horizon the outgoing flux at infinity 
looks like eq. (\ref{rsup}) (as in \cite{MuWiZe}) 
\be
\langle in| T_{uu}| in\rangle \to \frac{1}{768\pi M^2} (1-6)
\label{flso}
\ee
indicating that the black hole ``antievaporates'' absorbing energy from the 
vacuum. On the other hand, as $r\to 2M$ 
\be
\langle in | T_{vv} | in\rangle \to - \frac{1}{768\pi M^2} \ ,
\label{hara}
\ee
i.e. one has the usual negative energy inflow, 
which makes the interpretation even 
more puzzling. \par \noindent
Finally, it is worth noting that if we used the $\langle T_{ab}\rangle$
of eqs. (\ref{xxv}), the one constructed by integrating the conservation 
equations $\nabla_a \langle T^a_b\rangle =0$, we would obtain 
(see \cite{KuLiVa}) , 
instead 
of eq. (\ref{flso}) and in the same limit, 
\be
\langle T_{uu}\rangle \to \frac{1}{768\pi M^2}(1-3)
\label{xxxvi}
\ee 
which unfortunately does not improve the situation. 
\par \noindent
Concluding this section, we arrive at the unsatisfactory situation in which 
the ``sophisticated'' 2D theory described by $S_{aind}$ produces a 
$\langle T_{ab}\rangle $ for the Schwarzschild black hole which, apart from the
$|B\rangle$ state, not only is in qualitative disagreement with all 
that is known about the 4D $\langle T_{\mu\nu} \rangle$, but, even more
seriously, it is physically unacceptable. Its use in backreaction models 
is therefore highly questionable.
%%%%%%%%%%%%%%%%%%%%%%%%%%%%%%%%%%%%%%%%%%%%%%%%%%%%%%%%%%%%%%%%
\sect{An improved theory}
%%%%%%%%%%%%%%%%%%%%%%%%%%%%%%%%%%%%%%%%%%%%%%%%%
As already said, the conformal anomaly determines only the Weyl non invariant 
part of the effective action, namely  $S_{aind}$. The complete effective 
action should also contain a part invariant under Weyl transformations. 
The authors of Ref. \cite{MuWiZe} tried to calculate this part
perturbatively, since unlike $S_{aind}$ it cannot be computed
exactly. Using a simple classical approximation to the heat kernel 
they proposed to add to $S_{aind}$ the following nonlocal Weyl invariant term
of the Coleman-Weinberg type
\be
S_{wi}= \frac{1}{8\pi}\int d^2 x \sqrt{-g}
\left\{
-(\nabla\phi)^2 \frac{1}{\Box} R+\phi R+
\left( -\Box\phi +
(\nabla\phi)^2 \right) \left[ 1- 
\ln \frac{\left( -\Box\phi +
(\nabla\phi)^2 \right)}{\mu^2} \right] \right\},
\label{wein}
\ee
where $\mu$ is an arbitrary renormalization scale. 
One sees that the nonlocal term in eq. (\ref{wein}) cancels exactly the 
second nonlocal term in $S_{aind}$ (see eq. (\ref{xxvii}) ), leaving 
as unique nonlocal term the Polyakov one 
\be
S_{imp}=S_{aind} + S_{wi}= 
\frac{1}{8\pi}\int d^2 x \sqrt{-g}\{- \frac{1}{12}R     
\frac{1}{\Box}R+\left( -\Box\phi +
(\nabla\phi)^2 \right) \left[ 1- 
\ln \frac{\left( -\Box\phi +
(\nabla\phi)^2 \right)}{\mu^2} \right] \} \ .
\label{xxxvii}
\ee
At first sight, the advantage of this new formulation of the 2D theory
is clear: the second nonlocal term in $S_{aind}$, responsible for the 
appearance of the unphysical $-6$ in the Hawking flux (see eqs. (\ref{rsup})
and (\ref{flso}) ) has disappeared. This is the main argument used in 
\cite{MuWiZe} to show the accordance of this model with the 4D picture     
of Hawking black hole evaporation. 
The flux at infinity is now given by the Polyakov term as in the naive theory 
of sections 2 and 3, leading to the expected value 
 $\frac{1}{768\pi M^2}$. However, let us analyse in some detail the 
components of $\langle T_{ab}^{imp}\rangle$ in this theory. 
We have 
\ba
\langle T_{ab}^{imp} \rangle &=& \langle T_{ab}^P\rangle +
\frac{1}{4\pi} \Big\{
-\frac{g_{ab}}{2}\Big[ -\Box\phi + (\nabla\phi)^2
-(\nabla\phi)^2\ln \frac{\left(  -\Box\phi + (\nabla\phi)^2\right)}{\mu^2} 
\nonumber \\
&& - \nabla^{c}\phi\nabla_{c}\ln \frac{\left(  -\Box\phi +
(\nabla\phi)^2\right)}{\mu^2} \Big]
-\partial_{a}\phi\partial_{b}\phi \ln \frac{\left(  -\Box\phi +
(\nabla\phi)^2\right)}{\mu^2} \nonumber \\
&&-\frac{1}{2}\partial_{a}\phi\partial_{b}\ln \frac{\left(  -\Box\phi
+(\nabla\phi)^2\right)}{\mu^2} -\frac{1}{2}\partial_{b}\phi
\partial_{a}\ln \frac{\left( -\Box\phi +
(\nabla\phi)^2\right)}{\mu^2} \Big\} \ .
\label{xxxviii}
\ea
As usual, choosing a conformal frame $\{ x^{\pm} \}$ we find 
\ba
\langle T_{\pm\pm}^{imp}\rangle &=& \langle T_{\pm\pm}^P\rangle -
\frac{1}{4\pi}\left\{ \partial_{\pm}\phi \partial_{\pm}\phi \ln
\frac{\left(
-\Box\phi
+(\nabla\phi)^2\right)}{\mu^2} +\partial_{\pm}\phi\partial_{\pm}
 \ln \frac{\left(  -\Box\phi
+(\nabla\phi)^2\right)}{\mu^2} \right\}, \nonumber \\
\langle T_{+-}^{imp} \rangle &=& \langle T_{+-} \rangle \ ,
\label{xxxix}
\ea
where the r.h.s. of the second of eqs. 
(\ref{xxxix}) is still given by eq. (\ref{xxiv}) since
$S_{wi}$ does not alter, by construction, the 
trace anomaly. In the above 
\be
\left(  -\Box\phi
+(\nabla\phi)^2\right)=4e^{-2\rho}(\partial_+\partial_-\phi
-\partial_+\phi\partial_-\phi)\ .
\label{notaz}
\ee
From eqs. (\ref{xxxix}) we see that being the term in the curl brackets local, 
the difference of $\langle T_{ab}^{imp}\rangle$ between two states is simply
the Schwarzian derivative as in eqs. (\ref{cacar})-(\ref{def}). 
Inserting the Schwarzschild 
solution in eq. (\ref{xxxix}) we find ($f=1-2M/r$) 
\ba
\langle T_{uu}^{imp}\rangle &=&
 \langle T_{uu}^P \rangle + \frac{1}{16\pi}
\left[ -\frac{f^2}{r^2}\ln \frac{f^{'}}{\mu^2 r} + \frac{f^2}{r}
(\frac{f^{''}}{f^{'}}-\frac{1}{r}) \right] \ ,
\nonumber \\
\langle T_{vv}^{imp}\rangle &=&
 \langle T_{vv}^P \rangle + \frac{1}{16\pi}
\left[ -\frac{f^2}{r^2}\ln \frac{f^{'}}{\mu^2 r} + \frac{f^2}{r}
(\frac{f^{''}}{f^{'}}-\frac{1}{r}) \right] \ ,
\label{xxxx}
\ea
where a prime indicates derivative with respect to 
$r$ and $\langle T_{uu}^P\rangle$, $\langle T_{vv}^P\rangle$ are given in 
section 3 for the different states. At first sight the above 
expression seems reasonable, just local vacuum polorization added to the 
Polyakov term. 
Let us consider, however, the case $f=1$, i.e. Minkowski space-time. 
One immediately sees in eqs. (\ref{xxxx}) that the argument of the $\ln$
vanishes and $\langle T_{ab}^{imp}\rangle $ diverges. Therefore Minkowski 
vacuum is no longer a solution of the theory. The calculation in the shell 
collapse case of $\langle in |T_{ab}^{imp}|in\rangle$   
for $v<v_0$ (i.e. in the flat portion of the spacetime inside the shell) 
becomes meaningless in this context. This divergence is analogous to the 
infrared divergence of the Coulomb-Weinberg potential in the massless case.
\\
However, in addition to the Minkowski problem, we will have dangerous 
divergences of $\langle T_{ab}^{imp}\rangle$ for static spacetimes in 
regions where the surface gravity $f'$ vanishes. Non extreme 
Reissner-Nordstr\"om spacetime is one such example. The surface gravity
vanishes for $r=\frac{Q^2}{M}$ which lies between the inner and the outer 
horizon and there $\langle T_{ab}\rangle$ diverges. A similar situation 
happens for the Schwarzschild-de Sitter spacetime. 
All such features are not expected on physical ground and up to now there is 
no 4D evidence of such phenomena.    
%%%%%%%%%%%%%%%%%%%%%%%%%%%%%%%%%%%%%%%%%%%%%%%%%%%%%%%%%%%%%%%%
\sect{The four dimensional interpretation}
%%%%%%%%%%%%%%%%%%%%%%%%%%%%%%%%%%%%%%%%%%%%%%%%%
For the reasons previously explained (Minkowski as ground state) 
we prefer to come back to the action $S_{aind}$ and try to understand whether 
it is possible or not to extract physically sensible results. 
Let us consider the $\langle T_{ab}\rangle$ in eqs. (\ref{xxix}) and 
(\ref{xxiv}) and 
calculate its covariant divergence $\nabla_a \langle T^a_b \rangle$. 
As expected on the basis of the difference between eqs. (\ref{xxv}) 
and (\ref{xxix}) the result is nonzero and reads, in the conformal frame 
$\{  x^{\pm} \}$, 
\be
\nabla_{\mu} \langle T^{\mu}_{\pm}\rangle = \frac{N}{2\pi}\left(
2\rho \partial_{\pm}\phi\partial_{+}\partial_{-}\phi +
\partial_{\pm}\phi\partial_{\pm}\phi\partial_{\mp} \rho +\partial_{\pm}
\phi\partial_{\mp}\phi \partial_{\pm}\rho + \partial_{\pm}\phi
\partial_+\partial_-\rho \right) \ .
\label{xxxxi}
\ee
The r.h.s. of these equations are proportional to the quantum part 
(not to all!) of the equation of motion of $\phi$. We can in fact rewrite them 
in a more elegant form 
\be
\partial_{\mp}\langle T_{\pm\pm}\rangle
+\partial_{\pm}\langle T_{+-}\rangle  - \Gamma^{\pm}_{\pm\pm}
\langle T_{+-}\rangle + \partial_{\pm}\phi \frac{\delta
S_{aind}}{\delta\phi} =0 \ .
\label{xxxxii}
\ee
This relation has a general validity and applies for all theories
described by an action $S=S\left[ g_{\mu\nu}, \phi \right]$.
It shows, for instance, that the 2D conservation equations are automatically
satisfied by $\langle T_{ab}^P\rangle $ for the simple reason that the 
Polyakov action $S_P$ does not depend on $\phi$. 
For the other theories we are concerned with in this paper a similar 
result is no longer valid. 
\par \noindent
The situation seems therefore rather unsatisfactory: we started from a 4D 
classical theory, reduced it to 2D by assuming spherical symmetry 
and we are now left with a 2D effective theory 
where the basic ingredient, the matter energy momentum tensor, 
is not even conserved! Fortunately, there is no such a paradox as seen 
from the 4D point of view. Consider the 4D action $S^{(4)}_{aind}$
which, by dimensional reduction under spherical symmetry, gives 
$S_{aind}$. We can then define the 4D energy momentum tensor 
$\langle T_{\mu\nu}^{(4)}\rangle$ (see also \cite{MuWiZe}) 
\be
\langle T_{\mu\nu}^{(4)}\rangle = \frac{1}{\sqrt{-g^{(4)}}}
\frac{\delta S_{aind}^{(4)}}{\delta g^{\mu\nu}_{(4)} } \ .
\label{xxxxiii}
\ee
Under spherical symmetry these equations translate into the following 
definitions ($a,b=1,2$)
\ba 
\langle T_{ab}^{(4)}\rangle &=& \frac{\langle
T_{ab}^{(2)}\rangle }{4\pi e^{-2\phi}} \ , \nonumber \\
\langle T_{\theta\theta} \rangle &=& \frac {\langle T_{\phi\phi} \rangle}
{\sin^2\theta} = \frac{1}{8\pi \sqrt{-g^{(2)}}}\frac{\delta
S_{aind}}{\delta
\phi} \ ,
\label{xxxxiv}
\ea
where we have explicitly inserted the superscripts ${}^{(2)}$ and 
${}^{(4)}$ for clarification. These allow us to reinterpret eqs. (\ref{xxxxii})
as the conservation equations of the 4D stress tensor 
$\langle T_{\mu\nu}^{(4)}\rangle$, i.e. eqs. (\ref{xxxxii}) can be 
rewritten simply as 
\be
\nabla_{\mu}\langle T_{\nu}^{(4) \mu }\rangle =0 \ .
\label{xxxxv}
\ee
From the explicit form of $S_{aind}$ eq. (\ref{xxvii}) we obtain, 
in a conformal frame $\{ x^{\pm} \}$, 
\be
8\pi \langle T_{\theta\theta} \rangle = -\frac{1}{2\pi \sqrt{-g^{(2)}}}
\left[ 2\rho \partial_+\partial_-\phi + \partial_-\rho \partial_+\phi
+ \partial_+\rho\partial_-\phi + \partial_+\partial_-\rho \right] \ . 
\label{titete}
\ee
The above discussion and the 4D interpretation of the failure of the 
2D conservation equations can repeated step by step for the improved theory
of eq. (\ref{xxxvii}). In that case the angular component of 
$\langle T_{\mu\nu}^{(4)}\rangle$ is
\ba
8\pi \langle T_{\theta \theta} \rangle &=& 
-\frac{1}{4\pi  \sqrt{-g^{(2)}}}
\{ -2\partial_+\partial_-\phi [ -2\rho +\ln \frac{\left( 4(\partial_+
\partial_-\phi -\partial_+\phi \partial_-\phi)\right)}{\mu^2} ]
\nonumber \\
&&-\partial_+\phi [ -2\partial_-\rho +\partial_- \ln 
 \left( 4(\partial_+
\partial_-\phi -\partial_+\phi \partial_-\phi)\right) ]
-\partial_-\phi  [ -2\partial_+\rho 
+\partial_+ \ln 
 ( 4(\partial_+
\partial_-\phi 
\nonumber \\
&& -\partial_+\phi \partial_-\phi) ) ]
+ 2\partial_+\partial_-\rho - \partial_+\partial_- 
\ln \left( 4(\partial_+
\partial_-\phi -\partial_+\phi \partial_-\phi)\right) \} \ .
\label{titemu}
\ea
Let us finally write in full the action of the theories we have examined. 
The first model (anomaly induced) is described by 
\be
S=S_{g} + S_{aind},
\label{totac}
\ee
where 
\be
S_{g}=\frac{1}{2\pi }\int d^2 x \sqrt{-g}e^{-2\phi}[R + 
2(\nabla\phi)^2 +2 e^{2\phi} -2 \Lambda]
\label{grav}
\ee
($\Lambda$ is the 4D cosmological constant) and 
\be
S_{aind}=-\frac{1}{2\pi}\int d^2 x \sqrt{-g}\left[\frac{1}{48} R
\frac{1}{\Box}R
-\frac{1}{4}(\nabla\phi)^2 \frac{1}{\Box} R + \frac{1}{4}\phi R\right].
\label{pacca}
\ee 
The resulting field equations are 
\ba
2r\nabla_a\nabla_b r + g_{ab}(1- (\nabla r)^2 -2r\Box r +\frac{1}{2}
\Lambda r^2 ) &=& 2\pi \langle T_{ab} \rangle \ ,
\nonumber \\
r\Box r - \frac{1}{2}r^2 R  -\frac{\Lambda r^2}{2}= 
-4\pi^2  \langle T_{\theta\theta}\rangle  \ ,
\label{bacequ}
\ea 
where we have used $r\equiv e^{-\phi}$, $\langle T_{ab}\rangle$
is given in eqs. (\ref{xxviii}) and $\langle T_{\theta\theta}\rangle $ in 
(\ref{titete}). The improved theory of section 5 is described by 
\be
S=S_g +  S_{imp} ,
\label{tonu}
\ee
where 
\be
S_{imp}= \frac{1}{8\pi}\int d^2 x \sqrt{-g}\left\{- \frac{1}{12}R 
\frac{1}{\Box}R +
\left( -\Box\phi +
(\nabla\phi)^2 \right) \left[ 1- 
\ln \frac{\left( -\Box\phi +
(\nabla\phi)^2 \right)}{\mu^2} \right] \right\}
\label{rotto}
\ee
and the field equations are the same (with obvious substitution of the 
source terms with eqs. (\ref{xxxviii}) and (\ref{titemu}) ).
For the improved theory Minkowski, as we have said, is not a self 
consistent solution of the equations of motion. The l.h.s of 
eqs. (\ref{bacequ}) vanishes identically (for $\Lambda=0$) whereas
$\langle T_{\theta\theta}^{(4) imp}\rangle$ and 
$\langle T_{ab}^{imp}\rangle$ diverge, as can be seen explicitly in 
eqs. (\ref{titemu}) and  (\ref{xxxviii}). However, the improved 
theory has other interesting solutions. \\ 
The presence of a scale
in the theory depletes, in this case, Minkowski space-time of its central role 
in favour of other geometries. Let us consider de Sitter spacetime,
which is a classical solution of $S_g$ with $\Lambda \neq 0$.
One can then show, by fine tuning the arbitrary renormalization 
scale $\mu$ in eq. (\ref{xxxvii}) (i.e. $\mu^2=\frac{2}{3}\Lambda$ )
and the 2D cosmological constant (that can always be added to the 
Polyakov term $S_P$), that for the de Sitter spacetime  
\ba
\langle dS | T_{ab}^{imp} |dS \rangle &=& 0 \ , \nonumber \\
\langle dS | T^{imp}_{\theta\theta} | dS\rangle  &= & 0 \ ,
\label{desisp}
\ea
where $|dS\rangle$ means de Sitter invariant state, obtained by 
choosing $\{\tilde  x^{\pm} \}$ as Gibbons - Hawking null coordinates
\cite{GibbHawk}.
The de Sitter spacetime does not acquire, in the improved theory of eq. 
(\ref{tonu}), quantum corrections and is therefore a self consistent 
solution of the semiclassical equations. Despite this fact, we feel rather 
uneasy with the unphysical results that this improved theory predicts
for Minkowski space. The same can be said for Hawking black hole 
evaporation as described by $S_{aind}$. 
In the next section we shall outline how, in our opinion, an effective 
2D theory which can positively
deal with black hole evaporation should look like.
%%%%%%%%%%%%%%%%%%%%%%%%%%%%%%%%%%%%%%%%%%%%%%%%%%%%%%%%%%%%%%%%
\sect{The physical stress tensor: a proposal}
%%%%%%%%%%%%%%%%%%%%%%%%%%%%%%%%%%%%%%%%%%%%%%%%%%%%%%%%%%%%%%%    
The satisfactory interpretation of $\langle T_{ab}^{(2)}\rangle$
and $\delta S / \delta \phi $ as part of  
$\langle T_{\mu\nu}^{(4)}\rangle$  along with the conservation equations 
(\ref{xxxxv}) encourage us to adopt a 4D point of view. 
An ``acceptable'' 2D effective action deduced from the trace anomaly 
($S_{aind}$)  and additional Weyl invariant terms should reproduce
at least the qualitative features of $\langle T_{\mu\nu}^{(4)}\rangle$ 
for the Schwarzschild
space time. 
We stress that the comparison can only be qualitative since the 
exact analytic expression can of course not be met by a simple 
2D theory. In particular, 
the 4D anomalous trace  $\langle T_{\alpha}^{(4)\alpha}
\rangle$ is a local expression involving $R^{(4)}$, 
$R_{\mu\nu}^{(4)}$, $C_{\alpha\beta\gamma\delta}$ (see eq. (\ref{tial})):
a much more complicated expression than our 2D analogous eq. (\ref{xxiii}).
Neverthless, we require that some characteristic features of 
$\langle T_{\mu\nu}^{(4)} \rangle$ should be reproduced by an
``acceptable'' 2D theory, namely, in the spirit of the Wald's axioms,
\par \noindent
i) conservation equations $\nabla_{\mu}\langle T^{(4) \mu}_{\nu}\rangle
=0$;
\par \noindent
ii) vanishing of $\langle T^{(4)}_{\mu\nu}\rangle$ for Minkowski 
vacuum;
\par \noindent
iii) locality (by this we mean state independence) of the 4D trace 
$\langle T^{(4)\alpha}_{\alpha}\rangle$ .
\par \noindent
Using the definitions given in eqs. (\ref{xxxxiv}) we have
\be
\langle T^{(4)} \rangle = e^{2\phi}\left( \frac{\langle T^{(2)}\rangle}{4\pi}
+ 2\langle T_{\theta\theta} \rangle \right) \ .
\label{xxxxvi}
\ee
Since the 2D trace anomaly $\langle T^{(2)}\rangle$ given by eq. (\ref{xxiii})
is already local, one has to require, in order to satisfy iii),
that $\langle T_{\theta\theta}\rangle$ has to enjoy the same property. 
Let us consider the  $\langle T_{\theta\theta}\rangle$ given by $S_{aind}$,
namely eq. (\ref{titete}). Under the scaling $\{ x^{\pm} \} \to \{ 
\tilde x^{\pm} \}$ it transforms in an ``anomalous'' way, i.e. 
\be
\langle \tilde x^{\pm} | T_{\theta\theta} | \tilde x^{\pm} \rangle
= \langle  x^{\pm} | T_{\theta\theta} |  x^{\pm} \rangle
- \frac{e^{-2\rho}}{8\pi^2}\{ (\partial_+\partial_-\phi)
\ln F G + \frac{1}{2} \partial_-\phi \frac{G'}{G} + \frac{1}{2} 
\partial_+\phi \frac{F'}{F} \} \ .
\label{tean}
\ee
Because of the curl brackets term, $\langle T_{\theta\theta}\rangle$ 
and hence $\langle T^{(4)\alpha}_{\alpha}\rangle $ 
is state dependent, contrary to
our assumption iii). 
Note that $\langle T^{imp}_{\theta\theta}\rangle$ of eq. (\ref{titemu}) 
is state independent, however we do not consider it as a good starting 
point since it diverges for Minkowski spacetime, making the improved 
theory of section 5 incompatible with the requirement ii). 
\par \noindent
Our task will be to find a modified version of the stress tensor, call it
$\langle T_{\mu\nu}^{(4) new}\rangle$, which does indeed fullfill all our
three requirements. This tensor should in principle derive by an 
effective action $S^{new}$ which is obtained implementing, as in section 5,
$S_{aind}$ with Weyl invariant terms. We are not able to construct 
$S^{new}$ explicitly, but we shall give a sketch of how the new tensor
should look like. \\
Let us select a conformal frame $\{ x^{\pm}= u,v \} $ of reference which 
we will specify later. In an arbitrary conformal frame $\{ \tilde 
x^{\pm} = U,V \} $, related to the previous one by the functions $F$ and $G$ 
as in eqs (\ref{deg}), (\ref{def}), 
we have, in the state $| \tilde x^{\pm} \rangle$,
\be
\langle \tilde x^{\pm}| T_{\theta\theta}|\tilde x^{\pm}\rangle =
- \frac{e^{-2\tilde \rho}}{4\pi^2} \{ 
\tilde \rho\partial_U\partial_V \phi +
\frac{1}{2} \partial_U\phi \partial_V \tilde \rho + \frac{1}{2}
\partial_V\phi\partial_U\tilde\rho + \frac{1}{2} \partial_V\partial_U
\tilde \rho \} \ .
\label{theimp}
\ee
Now define 
\be
\langle \tilde x^{\pm}| T_{\theta\theta}^{new}|\tilde x^{\pm}\rangle =
\langle \tilde x^{\pm}| T_{\theta\theta}|\tilde x^{\pm}\rangle 
+\frac{e^{-2\tilde\rho}}{4\pi^2} \{ (\partial_U\partial_V\phi )\frac{1}{2}
\ln F G + \frac{1}{4} (\partial_V\phi)\frac{\dot F}{F} +
\frac{1}{4} (\partial_U\phi) \frac{\dot G}{G} \},
\label{gigio}
\ee
where $\dot F \equiv \frac{dF}{dU}$ and $\dot G \equiv \frac{dG}{dV}$.
Using eq. (\ref{tean}) and $\tilde \rho = \rho + \frac{1}{2} \ln F G $
one can show that
\be
\langle \tilde x^{\pm} | T_{\theta\theta}^{new}| \tilde x^{\pm}\rangle =
- \frac{ e^{-2\rho}}{4\pi^2} \{ (\partial_u\partial_v\phi)\rho + 
\frac{1}{2} \partial_u\phi \partial_v\rho + \frac{1}{2} 
\partial_v\phi \partial_u\rho +  \frac{1}{2}\partial_u\partial_v\rho
 \} = \langle x^{\pm} |
T_{\theta\theta} | x^{\pm}\rangle \ .
\label{toroi}
\ee
This means that in every state the expectation value of $\langle 
T_{\theta\theta}^{new} \rangle$ is given by $\langle x^{\pm} |
T_{\theta\theta} | x^{\pm}\rangle$. So we have achieved the state 
independence: under the conformal transformation \\
$\{ x^{\pm} \} \to 
\{ \tilde x^{\pm} \}$ $\langle T_{\theta\theta}\rangle$ remains 
unchanged as required by iii). 
The reference state $|x^{\pm}\rangle$ is chosen, in view of the requirement 
ii), such that $\{ x^{\pm} \}$ are Minkowskian coordinates at infinity.
For the Schwarzschild spacetime this implies that $|x^{\pm}\rangle =
|B\rangle $.  \\
Having defined $\langle T_{\theta\theta}^{new}\rangle$ which allows 
$\langle T^{(4) \alpha}_{\alpha}\rangle$ to be state independent, 
in order to enforce the conservation equations $\nabla_{\mu}\langle 
T^{(4)\mu}_{\nu}\rangle=0$ , as required by i),we have to redefine 
$\langle T_{\pm\pm}\rangle$ as well. 
In the $\{ \tilde x^{\pm}= U, V \}$ frame we then have
\ba
\langle \tilde x^{\pm} |
T_{UU}^{new} |\tilde x^{\pm}\rangle &=& \frac{1}{2\pi} 
\{ [ (\partial_U\phi)^2 \tilde \rho + \frac{1}{2} 
\partial_U \int dV \partial_U\phi\partial_V\phi ] 
-\frac{1}{4\pi} [ (\partial_U\phi)^2 \ln F G +
\nonumber \\ 
&&\dot F  \int dV
\partial_V\phi\partial_U\phi ] 
-\frac{1}{4\pi}(-2\partial_U\tilde \rho 
\partial_U\phi + \partial_U^2\phi) \} +
\langle \tilde x^{\pm} | T_{UU}^P|\tilde x^{\pm}\rangle 
\label{morri}
\ea
where the last term is given in detail in section 2. 
Similarly,
\ba
\langle \tilde x^{\pm} |
T_{VV}^{new} |\tilde x^{\pm}\rangle &=& \frac{1}{2\pi} 
\{ [ (\partial_V\phi)^2 \tilde \rho + \frac{1}{2} 
\partial_V \int dU \partial_U\phi\partial_V\phi ]            
-\frac{1}{4\pi} [ (\partial_V\phi)^2 \ln F G +
\nonumber \\ 
&& \dot G \int dU
\partial_V\phi\partial_U\phi ] 
-\frac{1}{4\pi}(-2\partial_V\tilde \rho 
\partial_V\phi + \partial_V^2\phi) \} +
\langle \tilde x^{\pm} | T_{VV}^P|\tilde x^{\pm}\rangle 
\label{macchu}
\ea
Note that under the conformal  transformation $\{ U,V\}\to \{ x^{\pm}=u,v\}$
the terms under curl brackets transform like a tensor, whereas the Polyakov 
term picks up the usual Schwarzian derivative. Summarizing, from 
eq. (\ref{morri}) and (\ref{macchu}) we have
\ba
\langle \tilde x^{\pm} | T_{uu}^{new} | \tilde x^{\pm} \rangle &=&
\frac{1}{2\pi} \left[ (\partial_u\phi)^2\rho + \frac{1}{2}
\partial_u \int dv \partial_u\phi\partial_v\phi ) \right]
- \frac{1}{4\pi} ( -2\partial_u\rho\partial_u\phi + \partial_u^2\phi)
\nonumber \\
&& + \frac{1}{12\pi} \left( \partial_u^2\rho -\partial_u\rho\partial_u\rho 
\right) 
+ \frac{1}{24\pi} \left( \frac{F''}{F} 
- \frac{1}{2} \frac{F^{'2}}{F^2} \right) 
\nonumber \\
&& = \langle x^{\pm} | T_{uu} | x^{\pm}\rangle + \frac{1}{24\pi} 
 \left( \frac{F''}{F} 
- \frac{1}{2} \frac{F^{'2}}{F^2} \right)
\label{picchu}
\ea
and similarly for $u$ interchanged with $v$ and $F$ with $G$. 
The 2D trace part remains unchanged 
\be
\langle \tilde x^{\pm} | T_{uv}^{new}|\tilde x^{\pm}\rangle =
\langle \tilde x^{\pm} | T_{uv}|\tilde x^{\pm}\rangle \ .
\label{ugual}
\ee
One can check that $\langle T_{\mu\nu}^{new}\rangle$ is conserved, has a 
trace which does not depend on the state and vanishes for 
Minkowski space-time, defined by $\rho=0$ and
$e^{-\phi}=(v-u)/2$: $  \langle M|T_{\mu\nu}^{new}|M\rangle=0 $.
\\
We should remind that our $\langle T_{\mu\nu}^{new} \rangle$ 
is defined modulo additional local terms which come from local Weyl 
invariant contributions that can be added to the 2D effective action.
These extra terms have to vanish in Minkowski space and being local
do not contribute to the Hawking radiation. 
\par \noindent
Let us now see
how our procedure works for the Schwarzschild black hole. 
Being the reference vacuum the Boulware one, 
$\langle B|T_{\mu\nu}^{new}|B\rangle$ has the same form as given in section 4,
in particular (omitting the superscript `new') 
\ba
  \langle B |T_{uu}| B\rangle & = &
\frac{1}{12\pi}\left(\frac{ff^{''}}{8}-\frac{f^{'2}}{16}\right)
+ \frac{1}{16\pi}\frac{f^2}{r^2}\ln f,
\nonumber \\
\langle B| T_{vv}| B\rangle & = &
\frac{1}{12\pi}\left(\frac{ff^{''}}{8}-\frac{f^{'2}}{16}\right)
+ \frac{1}{16\pi}\frac{f^2}{r^2}\ln f ,
\nonumber \\
\langle B| T_{uv}| B\rangle & = & \frac{1}{96\pi}ff^{''} +
\frac{1}{16\pi}\frac{ff^{'}}{r} \ ,
\label{xxx}
\ea
where $f=1-\frac{2M}{r}$. 
As expected, these expressions vanish for $M=0$, confirming that
Minkowski space-time is a solution of the backreaction equations. 
\\
In the Hartle-Hawking state we have 
\be
\langle H| T_{uu}|H\rangle (= \langle H| T_{vv}|H\rangle) 
= \langle B| T_{uu}|B\rangle  + \frac{1}{768\pi M^2} \ .
\label{hahane}
\ee
As $r\to\infty$ the first term on the r.h.s. of the above equation 
vanishes, confirming that $|H\rangle$ asymptotically describes
radiation in thermal equilibrium at the correct Hawking temperature 
$T_H$. 
Note that we have a logarithmic divergence on the horizon (in Kruskal 
coordinates). This is however integrable and does not affect the regularity 
of the semiclassical geometry. \\
Finally, for the dynamical situation of a black hole 
formed by the gravitational
collapse of a shock-wave at $v=v_0$ we have, from eq. (\ref{picchu}),
(for $v>v_0$) 
\ba
  \langle in|
T_{uu}| in \rangle &=&
\frac{1}{12\pi}\left(\frac{ff^{''}}{8}-\frac{f^{'2}}{16}
- \frac{3}{4}\frac{M^2}{r^4(u,v_0)} +
\frac{M}{2 r^3(u,v_0)}\right)
\nonumber \\
&&+ \frac{1}{16\pi} \left( \frac{f^2}{r^2}\ln f
- \frac{f^2(u,v_0)}{r^2(u,v_0)}    \right),
\nonumber \\
\langle in| T_{vv}| in \rangle &=&
\langle B| T_{vv}| B \rangle \ .
\label{lacca}
\ea
As the shell radius approaches the horizon $|in\rangle \to |U\rangle$ 
and we have 
\be
\langle U|T_{uu}| U\rangle =     
 \frac{1}{768\pi M^2} (1-\frac{2M}{r})^2
\left[ 1+ \frac{4M}{r} + \frac{12 M^2}{r^2} \right] 
\label{gippo}
\ee
leading to the expected flux at infinity 
\be
\langle T_{uu} \rangle \to \frac{1}{768\pi m^2}.
\label{giusto}
\ee
%%%%%%%%%%%%%%%%%%%%%%%%%%%%%%%%%%%%%%%%%%%%%%%%%%%%%%%%%%%%%%%%
\sect{Conclusions}
%%%%%%%%%%%%%%%%%%%%%%%%%%%%%%%%%%%%%%%%%%%%%%%%%%%%%%%%%%%%%%%%
The purpose of this paper was to extend the analysis of quantum black holes
from the framework of Polyakov theory to more appealing 
2D theories ($S_{aind}$, $S_{imp}$) whose link to the physical four 
dimensions appears more direct. 
Despite the appeal of these ``sophisticated theories'', their predictions 
turned out to be unacceptable : negative Hawking flux ($S_{aind}$) - 
nonzero (diverging!) renormalized stress tensor for Minkowski 
space-time ($S_{imp}$). \\
Given these astonishing results, we have attempted to modify the matter stress
energy tensor by imposing three requirements on it. The first two,
conservation equations and vanishing in Minkowski space, are quite obvious;
the third (state independence of the 4D trace) escapes from the strict 
two-dimensional point of view of all these models. 
However, as the discussion of the conservation equations has clearly 
shown, a correct handling and understanding of these theories can only be 
four dimensional. \\
Within our approach sensible results emerge that can be positively
compared to the 4D ones. 
We are well aware that our method may appear rather rough being not 
based on an elegant effective action like $S_{aind}$ and $S_{imp}$.
Unfortunately, as they stand   $S_{aind}$ and $S_{imp}$ 
cannot be the final answer. \\
One should however not exclude the possibility that there is no way 
of extracting sensible results from these hybrid lower dimensional 
theories and the only true improvement of the Polyakov theory
for the description of quantum black holes has to be genuinely 4D.

\vspace{1cm}
{\bf Acknowledgements}

A.F. wishes to thank R. Bousso for collaboration at an earlier stage
of this work and N. Kaloper, J. Rahmfeld for useful discussions.

\end{document}